\documentclass[sigconf]{acmart}
\acmConference[EASE 2026]{The 30th International Conference on Evaluation and Assessment in Software Engineering}{9–12 June, 2026}{Glasgow, Scotland, United Kingdom}
\usepackage{booktabs}
\usepackage{array}
\usepackage{pgfplots}
\pgfplotsset{compat=1.18}
\usepackage{tikz}
\usepackage{tcolorbox}
\usepackage{url}

\usepackage{listings}
\usepackage{xcolor}

\lstdefinelanguage{Rust}{
  keywords={fn, let, mut, String, from, println, vec, drop, push, usize},
  comment=[l]{//},
  morecomment=[s]{/*}{*/},
  morestring=[b]",
}

\newcommand{\revOne}[1]{\textcolor{black}{#1}}

\AtBeginDocument{%
  }

\begin{document}

\title{Mitigating False Positives in Static Memory Safety Analysis of Rust Programs via Reinforcement Learning}
%Leveraging Reinforcement Learning to Reduce False Positives in Static Memory Safety Analysis of Rust Programs}

\author{P Akilesh}
\affiliation{%
  \institution{Indian Institute of Technology}
  \city{Tirupati}
  %\state{Andhra Pradesh}
  \country{India}
}
\email{cs22b040@iittp.ac.in}

\author{Leuson Da Silva}
\affiliation{%
  \institution{Polytechnique Montreal}
  \city{Montreal}
  \country{Canada}}
\email{leuson-mario-pedro.da-silva@polymtl.ca}

\author{Foutse Khomh}
\affiliation{%
  \institution{Polytechnique Montreal}
  \city{Montreal}
  \country{Canada}}
\email{foutse.khomh@polymtl.ca}

\author{Sridhar Chimalakonda}
\affiliation{%
  \institution{Indian Institute of Technology}
  \city{Tirupati}
  %\state{Andhra Pradesh}
  \country{India}
}
\email{ch@iittp.ac.in}

\renewcommand{\shortauthors}{Akilesh et al.}
\newcommand{\ak}[1]{\textcolor{teal}{{\it [Akil: #1]}}}
\newcommand{\lm}[1]{\textcolor{purple}{{\it [Leuson: #1]}}}
\newcommand{\Foutse}[1]{\textcolor{red}{{\it [Foutse: #1]}}}
\begin{abstract}

Static analysis tools are essential for ensuring memory safety in Rust programs, particularly as Rust gains adoption in safety-critical domains. However, existing tools such as Rudra and MirChecker suffer from high false positive rates, which diminish developer trust, increase manual review effort, and may obscure genuine vulnerabilities. This paper presents a novel reinforcement learning (RL)-based approach for automatically classifying and suppressing spurious warnings in static memory safety analysis for Rust. To achieve this, we design an RL agent that learns a warning suppression policy by extracting contextual features from Rust’s Mid-level Intermediate Representation (MIR) and optimizing its decisions through interaction with static analysis outputs. To improve decision quality, we integrate dynamic validation via cargo-fuzz as an auxiliary feedback mechanism, allowing the agent to selectively validate suspicious warnings through targeted fuzz testing. Our evaluation shows that the proposed approach significantly outperforms state-of-the-art LLM-based baselines, achieving 65.2\% accuracy and an F1 score of 0.659, an improvement of 17.1\% over the best LLM baseline. With a recall of 74.6\%, our method successfully identifies nearly three-quarters of true bugs while substantially reducing false positives, improving precision from 25.6\% in raw Rudra output to 59.0\%. Incorporating dynamic fuzzing further boosts performance, yielding additional improvements of 10.7 percentage points in accuracy and 8.6 percentage points in F1 score over the RL-only variant. Overall, our work demonstrates that combining reinforcement learning with hybrid static–dynamic analysis can substantially reduce false positives and improve the practical usability of memory safety verification tools for Rust.
\end{abstract}



 %Reduction, Program Analysis, Fuzzing, MIR Analysis}

\maketitle

\section{Introduction}

The demand for memory-safe systems programming has catalyzed the rapid adoption of Rust across critical domains, including operating systems, cloud infrastructure, embedded systems, and security-sensitive applications \cite{sharma2024rust}. 
Major technology organizations have embraced Rust for its unique combination of performance and safety: the Linux kernel officially integrated Rust support in version 6.1~\cite{torvalds2022rust}, Microsoft has adopted Rust for Windows components~\cite{microsoft2019rust}, and companies like AWS, Cloudflare, and Discord report significant reliability improvements after migrating performance-critical services from C/C++ to Rust~\cite{aws2020rust, cloudflare2020rust}. 
Such widespread adoption stems from Rust's ownership model and type system, which provide compile-time guarantees that eliminate entire classes of memory safety vulnerabilities, including use-after-free, double-free, and data races, without requiring garbage collection~\cite{jung2017rustbelt, matsakis2014rust}.

Despite these powerful safety guarantees, the reality of systems programming necessitates escape hatches. 
Rust's \texttt{unsafe} keyword permits operations that the compiler cannot statically verify. %: dereferencing raw pointers, calling foreign functions, accessing mutable static variables, implementing unsafe traits, and manipulating union fields. 
Empirical studies reveal that unsafe code is pervasive throughout the Rust ecosystem, with 25--30\% of packages on crates.io containing at least one unsafe block~\cite{evans2020unsafe, astrauskas2020unsafe}. 
% Even the Rust standard library relies extensively on unsafe code to implement fundamental abstractions like \texttt{Vec}, \texttt{Arc}, and \texttt{Mutex}. This prevalence is not merely convenience; unsafe code is essential for foreign function interfaces, performance-critical optimizations, and low-level data structure implementations that cannot be expressed within Rust's safe subset~\cite{qin2020understanding}. Critically, memory safety violations in Rust programs invariably originate from incorrect unsafe code or its encapsulation~\cite{xu2021memory}.
Even the Rust standard library relies on unsafe code for core abstractions like \texttt{Vec} and \texttt{Mutex}, as it is essential for FFI, performance-critical optimizations, and low-level implementations that cannot be expressed in safe Rust~\cite{qin2020understanding}.

To address the verification challenges introduced by unsafe code, previous studies have developed static analysis tools based on Rust’s Mid-level Intermediate Representation (MIR) to detect memory safety violations at scale. 
% Representative examples include Rudra~\cite{bae2021rudra}, which performs ecosystem-wide analysis, uncovering over 264 previously unknown bugs; MirChecker~\cite{li2021mirchecker}, which applies symbolic execution on MIR to detect panics, memory leaks, and concurrency issues; and Yuga~\cite{ghimire2022yuga}, which targets lifetime and borrowing violations via interprocedural dataflow analysis. 
Representative MIR-based tools include Rudra~\cite{bae2021rudra}, MirChecker~\cite{li2021mirchecker}, and Yuga~\cite{ghimire2022yuga}.
Despite this success, false positives remain the dominant barrier to practical adoption, as current SOTA tools exhibit severe imprecision.
% As reported in Figure~\ref{fig:fp_rates} MirChecker reports false positives in 95.1\% of warnings, Rudra in 50\%, and Yuga in 46.15\%. ~\cite{bae2021rudra, ghimire2022yuga, li2021mirchecker}. 
Prior studies report that MirChecker produces false positives in 95.1\% of its warnings, while Rudra and Yuga exhibit substantially lower false-positive rates of 50\% and 46.15\%, respectively \cite{bae2021rudra, ghimire2022yuga, li2021mirchecker}.
% \ak{Since we removed the figure, we can cite it from the paper}
Such rates impose substantial manual review effort and lead to warning fatigue, where developers increasingly distrust or ignore analysis results~\cite{christakis2016developers, johnson2013static}. 
Meanwhile, empirical studies show that false positive rates above 20–30\% frequently result in tool abandonment in industrial settings~\cite{bessey2010few, sadowski2018lessons}. 
% As a result, many MIR-based analyzers, while powerful, remain impractical for routine use in safety-critical development.

%The false positive problem in static analysis reflects a fundamental tension between soundness and precision. Sound analysis tools aim to report all potential bugs (avoiding false negatives) but often sacrifice precision through conservative approximations of program behavior, particularly when reasoning about complex control flow, aliasing relationships, and higher-order functions~\cite{cousot1977abstract}. 
To overcome these weaknesses, different approaches have been explored. 
For example, traditional approaches have relied on manual tuning of analysis heuristics, pattern-based filtering rules derived from observed false positive patterns, or developer-provided annotations to guide analysis~\cite{ayewah2008evaluating, kremenek2004finding}. 
While effective in limited contexts, these techniques require significant domain expertise, lack adaptability to evolving codebases, and fail to generalize across different projects or programming idioms. 
More recent efforts have explored machine learning techniques for warning prioritization and classification~\cite{heckman2008establishing, kim2007warnings, ruthruff2008predicting}.
These approaches typically depend on hand-crafted features, supervised learning with limited labeled training data, or models that cannot adapt post-deployment as code patterns evolve.

To address these challenges, we introduce a reinforcement learning–based approach that automatically learns policies for classifying and suppressing false positive warnings in Rust static analysis. 
Unlike supervised learning methods that rely on large labeled datasets and remain fixed after training, 
% reinforcement learning allows an agent to optimize suppression decisions through interaction with analysis outputs, balancing false positive reduction against preserving true vulnerability detection~\cite{sutton2018reinforcement}. 
reinforcement learning provides a general framework for optimizing sequential decision policies through interaction with an environment and reward feedback~\cite{sutton2018reinforcement}. In our setting, this enables an agent to learn suppression strategies that trade off false-positive reduction against preserving true vulnerability detection.

Our approach leverages rich semantic features extracted directly from Rust’s MIR, enabling precise reasoning about control flow, type and lifetime constraints, and unsafe operations, capabilities that are difficult to achieve at the source level, particularly in the presence of generics and macro expansion. 
We instantiate our approach on top of Rudra, a MIR-based analyzer, whose scope and warning diversity make it well-suited for learning-based suppression \cite{bae2021rudra}.
To further improve decision confidence, the agent selectively integrates dynamic validation via cargo-fuzz when static evidence is insufficient, gathering concrete runtime feedback to resolve ambiguous warnings~\cite{bohme2017directed, klees2018evaluating}. 
This hybrid static–dynamic formulation enables cost-aware suppression policies that strategically allocate expensive fuzzing resources, bridging the gap between conservative static analysis and concrete program behavior while preserving scalability.

We evaluate our approach through a large-scale empirical study. 
We first modernized the Rudra static analyzer to support the current Rust ecosystem and applied it to approximately 20,000 crates from crates.io, producing a curated dataset of 4,879 warnings spanning diverse unsafe usage patterns. 
Ground truth labels are established via systematic manual classification by domain experts in Rust memory safety. 
We then benchmark our RL-based approach against state-of-the-art large language models, including CodeLlama 34B, Llama3 (8B and 70B), Mixtral 8×7B, ChatGPT-4o mini, and Claude Opus 4.1, representing current LLM capabilities for code analysis~\cite{chen2021evaluating, roziere2023code,grattafiori2024llama3herdmodels,jiang2024mixtralexperts}. 
Our results show that the RL agent with fuzzing integration achieves 65.2\% accuracy and a 0.659 F1 score, outperforming the best LLM baseline by 17.1 percentage points. 
With a recall of 74.6\%, our approach identifies nearly three-quarters of true bugs while more than doubling precision compared to Rudra’s raw output (from 25.6\% to 59.0\%). 
Selective dynamic fuzzing provides additional gains, improving accuracy by 10.7 percentage points and F1 score by 8.6 percentage points over the RL-only variant.

\begin{comment}
\begin{enumerate}
\item A reinforcement learning framework for false positive reduction in static memory safety analysis, capable of learning suppression policies through interaction with analysis outputs and adapting to diverse code patterns.

\item A hybrid static-dynamic approach that integrates cargo-fuzz as an RL action, enabling targeted dynamic validation to resolve ambiguous warnings with high precision.

\item A large-scale curated dataset of 4,879 manually-labeled static analysis warnings from 20,000 Rust crates, providing a valuable resource for future research in program analysis and machine learning for software engineering.

\item Comprehensive empirical evaluation demonstrating that our RL-based approach outperforms state-of-the-art LLM baselines in classification accuracy while maintaining practical scalability for real-world deployment.

\item An updated implementation of Rudra compatible with the modern Rust ecosystem, enabling continued research and practical application of static analysis for memory safety verification.
\end{enumerate}
\end{comment}

Overall, the primary contributions of this work are as follows:
\begin{enumerate}
\item A reinforcement learning–based framework for reducing false positives in Rust static memory safety analysis %by learning suppression policies from analysis feedback 
\cite{online_appendix}.

\item A hybrid static–dynamic formulation that incorporates cargo-fuzz as a selective RL action. % to validate ambiguous warnings.

\item A curated dataset of 4,879 manually labeled static analysis warnings collected from approximately 20,000 Rust crates.

\item A large-scale empirical evaluation showing that our approach outperforms state-of-the-art LLM-based baselines while remaining practical for real-world use.

\item A modernized implementation of Rudra compatible with the current Rust ecosystem \cite{rudra_github}.
\end{enumerate}

\section{Background}
\label{sec:background}

This section provides foundational concepts for understanding our reinforcement learning-based approach. 
First, we discuss Rust's memory safety model, followed by static analysis for Rust, and the adoption of fuzzing. 
% \lm{If we need space, we can remove the discussion of RL.}

\subsection{Rust's Memory Safety Model}

Rust is a systems programming language that provides memory safety without garbage collection through a sophisticated type system and ownership model~\cite{matsakis2014rust}. 
Unlike C/C++, Rust enforces safety at compile time, eliminating entire classes of memory errors.
Rust's safety guarantees rest on three fundamental principles:

\textbf{Ownership.} Every value in Rust has a single owner that statically determines its lifetime. 
When the owner goes out of scope, the value is automatically deallocated through RAII-style drop semantics~\cite{jung2017rustbelt}. 
This prevents use-after-free and double-free errors in safe Rust code (see Listing \ref{list1}):

\begin{lstlisting}[language=Rust, caption={Ownership example showing move semantics}, label=list1]
fn ownership_example() {
    let s1 = String::from("hello");
    let s2 = s1; // s1 moved to s2
    // println!("{}", s1); // Error: s1 no longer valid
    println!("{}", s2); // OK
} // s2 dropped here
\end{lstlisting}

\textbf{Borrowing.} Values can be temporarily borrowed through references without transferring ownership. 
The borrow checker ensures that references do not outlive the data they point to~\cite{matsakis2014rust} (see Listing \ref{list2}):

\begin{lstlisting}[language=Rust, caption={Borrowing example with reference passing}, label=list2]
fn borrowing_example() {
    let s = String::from("hello");
    let len = calculate_length(&s); // Borrowed
    println!("{} has length {}", s, len); 
}
fn calculate_length(s: &String) -> usize {
    s.len()
} // s goes out of scope (only borrowed)
\end{lstlisting}

\textbf{Aliasing XOR Mutability.} At any time, a value may have either multiple shared references or one mutable reference, but never both~\cite{jung2017rustbelt}. 
This prevents data races and iterator invalidation (see Listing \ref{list3}):

\begin{lstlisting}[language=Rust, caption={Aliasing XOR mutability enforcement}, label=list3]
fn aliasing_xor_mutability() {
    let mut data = vec![1, 2, 3];
    let r1 = &data; // Shared reference
    let r2 = &data; // OK: multiple shared refs
    // let r3 = &mut data; // Error: cannot borrow as mutable
    drop(r1); drop(r2); // End shared borrows
    let r3 = &mut data; // OK: can mutably borrow
    r3.push(4);
}
\end{lstlisting}

Despite Rust's strong safety guarantees, certain operations essential to systems programming cannot be verified by the compiler~\cite{astrauskas2020unsafe}. 
To support these cases, Rust provides the \texttt{unsafe} keyword, which permits five specific capabilities: dereferencing raw pointers, calling unsafe functions, accessing mutable static variables, implementing unsafe traits, and accessing union fields.

%Unsafe code is pervasive in practice, as empirical studies show that 25--30\% of crates.io packages contain unsafe blocks~\cite{evans2020unsafe, astrauskas2020unsafe}. Even the Rust standard library relies extensively on unsafe code to implement core abstractions like \texttt{Vec}, \texttt{Arc}, and \texttt{Mutex}. Critically, when unsafe code is encapsulated within safe APIs, the API author assumes responsibility for ensuring that no sequence of valid inputs can trigger undefined behavior. As a result, memory safety violations in Rust programs originate from unsafe code itself, rather than from its safe callers \cite{qin2020understanding}. 
%This makes unsafe code---not its callers---solely responsible for safety violations~\cite{qin2020understanding}.

\subsection{Static Analysis for Rust}

\begin{comment}
Static analysis tools leverage Rust’s intermediate representations to detect memory safety violations at scale. 
While different tools have been proposed and developed \cite{li2021mirchecker,cui2023safedrop,carter2016smack,ghimire2022yuga}, we focus on Rudra~\cite{bae2021rudra}, the primary tool analyzed here, due to its ecosystem-scale scope, focus on unsafe code, and ability to generate a large and diverse set of warnings suitable for learning-based false positive reduction.
In its original evaluation, Rudra analyzed over 43,000 packages, discovering 264 previously unknown bugs, including multiple CVEs in the Rust standard library. 
However, it suffers from high false-positive rates, highlighting the need for automated false-positive reduction.
\end{comment}
% \revOne{In this work, we focus on Rudra~\cite{bae2021rudra}, which analyzed over 43,000 packages, discovering 264 previously unknown bugs, but suffers from high false-positive rates, highlighting the need for automated reduction.}

\revOne{Static analysis tools leverage Rust's intermediate representations (IRs) to detect memory safety violations at scale. In this work, we focus on Rudra~\cite{bae2021rudra}, which analyzed over 43,000 packages, discovering 264 previously unknown bugs, but suffers from high false-positive rates, highlighting the need for automated reduction.}
%\subsubsection{The Rudra Static Analyzer} 
Rudra performs ecosystem-scale analysis to detect memory safety bugs in unsafe Rust code. 
Implemented as a custom compiler driver, Rudra intercepts compilation after type checking to inject analysis algorithms. 
Its key innovation is hybrid IR analysis, which combines High-Level IR (HIR), preserving source-level structure, with Mid-Level IR (MIR), which exposes explicit control flow. 
This design enables generic-aware analysis that is difficult to achieve at the LLVM IR level. 

\begin{lstlisting}[language=Rust, caption={HIR vs.\ MIR 
representation of a simple ownership transfer.}, 
label={lst:hir-mir}]
// Source / HIR-level (preserves structure):
let s1 = String::from("hello");
let s2 = s1; // ownership moved

// MIR-level (explicit move + invalidation):
_1 = String::from(const "hello"); // s1
_2 = move _1;                     // s2
// _1 is now invalid (move tracked explicitly)
\end{lstlisting}

\revOne{As shown in Listing~4, MIR makes ownership transfers explicit as \texttt{move} operations, allowing Rudra to precisely track lifetime-bypassing flows that are implicit at the HIR level.}

Rudra implements two primary algorithms. 
The \textbf{Unsafe Dataflow (UD)} checker performs taint-style tracking to identify flows from lifetime-bypassing operations (e.g., uninitialized memory and duplicated values) to potentially dangerous operations. 
It approximates dangerous points using unresolvable generic functions, functions whose implementation depends on caller-supplied type parameters, leading to conservative over-approximation and false positives. 
The \textbf{Send/Sync Variance (SV)} checker analyzes thread-safety properties of generic types by inferring required trait bounds from public APIs and comparing them against concrete implementations.

\begin{comment}    
\subsubsection{Other Rust Static Analyzers} 
MirChecker~\cite{li2021mirchecker} applies symbolic execution on MIR to detect runtime panics, memory leaks, and concurrency issues, but reports even higher false positive rates (95.1\%). SafeDrop~\cite{cui2023safedrop} focuses on memory deallocation bugs through data-flow analysis. Yuga~\cite{ghimire2022yuga} targets lifetime and borrowing violation through interprocedural analysis. In contrast, tools such as SMACK~\cite{carter2016smack} and Prusti~\cite{astrauskas2019prusti} provide formal verification guarantees but require manual annotations and do not scale to ecosystem-wide analysis.
\end{comment}

%\subsubsection{Bug Patterns in Unsafe Rust}

Rudra targets three critical classes of bugs that arise from incorrect use of unsafe code in Rust~\cite{bae2021rudra, xu2021memory}.
First, when Rust panics, stack unwinding executes destructors as control propagates upward, leading to \textbf{Panic Safety} issues.
If unsafe code temporarily violates internal invariants (e.g., mutating vector length during reorganization), a panic triggered by a caller-provided closure can cause use-after-free or double-free.
A representative example is CVE-2020-36317 in \texttt{String::retain()}, where panicking closures left strings in an invalid UTF-8 state \cite{cve-2020-36317}.
%\footnote{\href{https://nvd.nist.gov/vuln/detail/cve-2020-36317}{CVE-2020-36317}}

\begin{lstlisting}[language=Rust, caption={Panic safety 
violation: unsafe invariant broken before panic.}, 
label={lst:panic-safety}]
fn unsafe_retain(v: &mut Vec<u8>) {
    let len = v.len();
    unsafe { v.set_len(0); } // invariant broken
    // if closure panics here, destructor runs
    // on invalid state -> use-after-free
    v.retain(|x| *x > 0);
    unsafe { v.set_len(len); }
}
\end{lstlisting}

\revOne{As shown in Listing~5, Rudra's UD checker flags the \texttt{set\_len} call as a lifetime-bypassing operation whose invariant may not be restored if a panic occurs before the matching restore.}

Second, \textbf{Higher-Order Invariants} are related to unsafe code that often relies on semantic assumptions about generic functions that are not captured by type signatures, such as purity or consistency of results~\cite{qin2020understanding}. 
Violations of these assumptions can lead to memory corruption.
For example, CVE-2020-36323 in \texttt{join()} assumed that \texttt{Borrow} implementations return consistent results; impure implementations violated this invariant, causing buffer overflows \cite{cve-2020-36323}. %\footnote{\href{https://nvd.nist.gov/vuln/detail/cve-2020-36323}{CVE-2020-36323}}
Finally, generic types must correctly propagate thread-safety requirements through appropriate \texttt{Send} and \texttt{Sync} bounds. 
Missing or incorrect bounds can allow data races through safe code. 
This \textbf{Send/Sync Variance} issue can be observed in CVE-2020-35905 in the \texttt{futures-rs} crate, which occurred because \texttt{MappedMutexGuard} failed to enforce bounds on mapped types, \revOne{enabling data races involving \texttt{Rc} (a single-threaded reference-counted smart pointer that is not thread-safe) \cite{cve-2020-35905}.}

\subsection{Fuzzing and Dynamic Analysis}

Fuzzing automatically generates test inputs to discover bugs through concrete execution~\cite{klees2018evaluating}. 
Modern coverage-guided fuzzers, such as AFL and libFuzzer, leverage execution feedback to efficiently explore program paths~\cite{zalewski2014american}. 
When combined with sanitizers, like AddressSanitizer, MemorySanitizer, and ThreadSanitizer, fuzzing enables high-confidence detection of memory safety and concurrency errors. 
Cargo-fuzz integrates libFuzzer into the Rust ecosystem~\cite{cargofuzz}. 
While effective for bugs with well-defined input surfaces, fuzzing faces challenges in exercising subtle semantic violations, such as Send/Sync errors that require specific thread interleaving or execution contexts. 
Moreover, due to its computational cost, indiscriminate adoption of fuzzing across all potential warnings is impractical at scale~\cite{klees2018evaluating}.

Static and dynamic analysis exhibit complementary trade-offs: static analysis offers broad coverage and generic reasoning, but often produces false positives, whereas dynamic analysis only explores executed paths but provides concrete evidence of real bugs~\cite{li2019hybrid}. 
Hybrid approaches combining both techniques have been shown to achieve superior results~\cite{bohme2017directed}. 
Our RL framework operationalizes this paradigm by learning when to apply static or dynamic validation based on their relative effectiveness for different classes of warnings.  

\section{Methodology and Approach}
\label{sec:methodology}

This section describes our reinforcement learning-based approach for addressing the occurrence of false positives in Rudra's static memory safety analysis. 
First, we begin by outlining the dataset construction process, followed by our feature extraction strategy, the RL formulation, and the integration of cargo-fuzz as a dynamic validation mechanism.

\subsection{Dataset Construction and Labeling}
\label{sec:dataset}

Our work begins with creating a large-scale dataset of Rudra warnings from real-world Rust code. 
To enable this, we first updated the Rudra static analyzer to be compatible with contemporary Rust compiler versions. 
% The original Rudra implementation was developed against an older Rust nightly compiler from the 2020–2021 timeframe and relies on MIR and compiler APIs that have since evolved.
% \lm{Please, double check the discussion provided here; I did based on the original paper of Rudra. However, if you know the specific versions of Rust, you can inform them in the text.}
% \ak{I have added the specific version of compiler}
The original implementation of Rudra was developed for the Rust nightly-2021-10-21 compiler \cite{rust1560} %\footnote{\href{https://doc.rust-lang.org/beta/releases.html\#version-1560-2021-10-21}{Version 1.56.0}} 
and depends on MIR representations and compiler APIs that have since evolved. As a result, it is incompatible with the current Rust toolchain. 
Our updated implementation targets the Rust nightly-2025-06-26 compiler \cite{rust1880} 
%\footnote{\href{https://doc.rust-lang.org/beta/releases.html\#version-1880-2025-06-26}{Version 1.88.0}} %\lm{Please, double check the links here.} 
and incorporates adaptations to modern MIR representations and changes to Rust’s type system introduced in later compiler releases. Our implementation is available online  \cite{rudra_github}. 
To ensure correctness, we validated our implementation against Rudra’s official regression test suite, which covers Unsafe Dataflow, Panic Safety, and Send/Sync Variance analyses with both positive and negative cases. We ran the test suite under the Rust nightly-2025-06-26 compiler, and all tests successfully passed (23/23 standard cases, 1/1 false-negative, and 3/3 false-positive cases), showing that our implementation correctly reproduces the behavior of the original analyzer.
% \Foutse{where is this update available? Can we point to it? Did we test that it is sound? How? Can we say a bit more about that process?}

%To ensure that these updates preserve the correctness and expected behavior of the original analyzer, we validated our implementation against the official Rudra regression test suite provided by the original repository. 
%This suite consists of curated Rust programs covering Unsafe Dataflow, Panic Safety, and Send/Sync Variance analyses, and includes both positive and negative test cases. 
%We executed the test harness using the \texttt{test.py} script included in the repository under the Rust nightly-2025-06-26 compiler. 
%All tests completed successfully: 23 out of 23 normal test cases passed, together with all false-negative (1/1) and false-positive (3/3) benchmark cases. 
%These results indicate that our updated implementation faithfully reproduces the behavior of the original Rudra analyzer on its official validation suite, providing confidence that the port to the modern Rust compiler preserves analyzer soundness and functionality.

Using the updated analyzer, we ran Rudra on approximately 20,000 crates from crates.io that contained at least one unsafe block, ensuring relevance to our analysis target.
This analysis produced 4,879 unique warnings generated by Rudra’s Unsafe Dataflow and Send/Sync Variance checkers. 
Each warning is associated with its source code location, the detected bug pattern (panic safety, higher-order invariant, or Send/Sync variance), the precision level reported by Rudra, and contextual information describing the surrounding code structure.
% \lm{I miss further details about it. Maybe, you provide further information later in this section.} \Foutse{yes, please provide precise information! We could even add an illustrative example!}
To illustrate the collected warnings, Listing~\ref{lst:rudra-aarc-json} shows the raw JSON output produced by Rudra for a warning in the \texttt{aarc} crate (v0.3.2) \cite{aarc032}.
%\footnote{\href{https://crates.io/crates/aarc/0.3.2}{Version 0.3.2}} 
The report includes the warning level, the responsible analyzer, a textual description, the precise source location, and the extracted code snippet.

%"start_line": 118,
%"start_col": 1,
%"end_line": 118,
%"end_col": 33,
%"code_snippet": "impl<T: 'static> Drop for Arc<T> {\n fn drop(&mut self) {\n let birth_era = unsafe { self.ptr.as_ref().birth_era };\n retire(self.ptr.cast(), decrement_strong_count::<T>, birth_era);\n }\n }"

% Listing~\ref{lst:rudra-aarc-code} presents the corresponding Rust source code extracted from the reported location.

% \begin{lstlisting}[language=Rust, caption={Unsafe destructor code flagged by Rudra in \texttt{aarc}.}, label={lst:rudra-aarc-code}]
% impl<T: 'static> Drop for Arc<T> {
%     fn drop(&mut self) {
%         let birth_era = unsafe { self.ptr.as_ref().birth_era };
%         retire(self.ptr.cast(), decrement_strong_count::<T>, birth_era);
%     }
% }
% \end{lstlisting}

%A key challenge in supervised learning approaches to false positive reduction is the availability of accurate ground truth labels. 
For each collected warning, one reviewer, with experience in Rust memory safety semantics, %\Foutse{who ix 'We'? Please be precise!!!} 
manually classified all 4,879 warnings through careful code review. %\Foutse{who are these reviewers? How many people participated?}
For each warning, the reviewer examined the surrounding code context, inspected the data-flow paths reported by Rudra, and determined whether the warning corresponded to a genuine soundness violation or a false alarm. 

\begin{lstlisting}[
    language=Rust, 
    stringstyle=\color{blue},
    caption={Rudra JSON report for an unsafe destructor in \texttt{aarc}.}, 
    label={lst:rudra-aarc-json}
]
{
    "level": "Warning",
    "analyzer": "UnsafeDestructor",
    "op_type": null,
    "description":"unsafe block detected in drop",
    "file": "aarc-0.3.2/src/smart_ptrs.rs",
    "start_line": 118, "start_col": 1, 
    "end_line": 118, "end_col": 33,
    "code_snippet": "impl<T: 'static> Drop for Arc<T> {...} }"
}
\end{lstlisting}

To assess labeling reliability, a second reviewer independently classified a random subsample of 150 warnings, yielding 82.7\% raw agreement (Cohen's $\kappa$ = 0.63), indicating substantial inter-rater agreement.
The labeling process focused on the underlying bug pattern rather than exploitability, as our goal is to improve the precision of static analysis rather than assess security impact. 
For example, in Listing~\ref{lst:rudra-aarc-json}, Rudra’s \texttt{UnsafeDestructor} analyzer reports a dereference inside a \texttt{Drop} implementation. These operations in destructors are known to be error-prone, as invalid pointer dereferences or concurrency violations at this stage may lead to undefined behavior. Finally, Rudra correctly classifies this instance as a potential higher-order invariant violation (true positive).

%For example, for the warning reported in Listing \ref{lst:rudra-aarc-json}, Rudra’s \texttt{UnsafeDestructor} analyzer reports an unsafe dereference inside a \texttt{Drop} implementation. 
%The JSON output provides structured metadata used in our dataset construction, while the Rust listing highlights the underlying code pattern. 
%Such unsafe operations in destructors are known to be error-prone, as invalid pointer dereferences or concurrency violations at this stage may lead to undefined behavior. 
%Accordingly, Rudra classifies this instance as a potential higher-order invariant violation. 
Overall, this process identified 1,247 true positives (25.6\%) and 3,632 false positives (74.4\%), available in our online appendix \cite{online_appendix}, highlighting the substantial false positive burden that motivates this work. We partitioned the dataset into training (70\%), validation (15\%), and test (15\%) sets using stratified sampling to preserve class distribution, yielding 732 warnings in the held-out test set used for final evaluation.

%\Foutse{who is 'we'? We should be precise!}

\begin{figure}[t]
\centering
\includegraphics[width=0.48\textwidth]{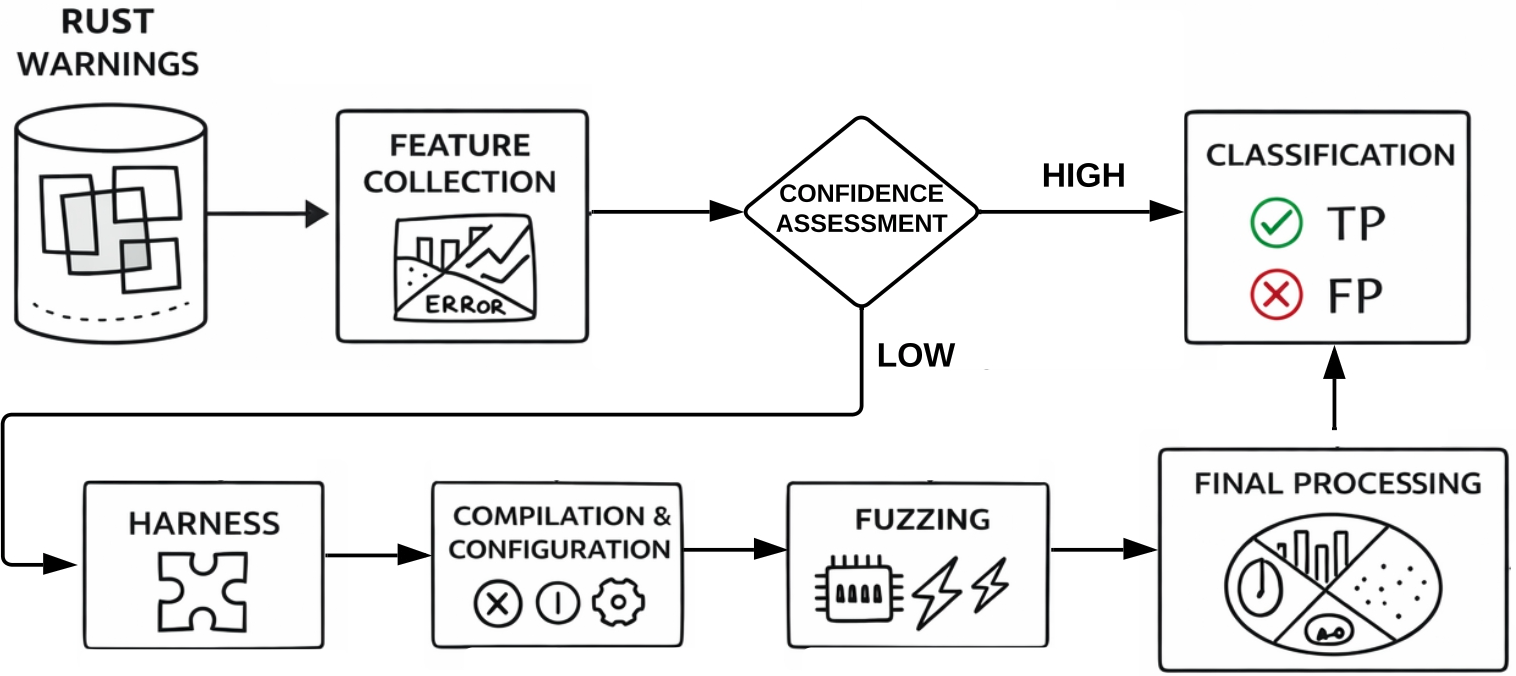}
% \caption{Overview of the RL-based fuzzing integration pipeline. The RL agent observes feature vectors and assesses confidence. For low-confidence warnings, it invokes targeted fuzzing through automated harness generation, compilation with sanitizers, and execution via cargo-fuzz. Fuzzing outcomes (bug found, clean execution, or inconclusive) are processed and used to augment the state vector, enabling the agent to make a final informed classification decision.}
\caption{Overview of the warning classification pipeline. Rust warnings are processed through feature collection and confidence assessment. High-confidence warnings are directly classified as true positives or false positives. Low-confidence warnings trigger automated harness generation, compilation and configuration, and fuzzing. The outcomes of fuzzing are aggregated in the final processing and used to make the final classification decision.}
\label{fig:rl_fuzzing_pipeline}
\end{figure}

\subsection{Feature Extraction from MIR}

Effective classification requires features that capture characteristics distinguishing true bugs (true positives) from false alarms (false positives). To this end, we extract features at multiple levels of abstraction, leveraging Rust’s Mid-level Intermediate Representation (MIR) to access rich semantic information while preserving awareness of generic types.

Our feature extraction process is organized into three primary categories. 
First, we extract MIR-level semantic features that characterize program behavior surrounding each warning. 
These include type system properties (e.g., number of generic parameters, presence of trait bounds, and generic nesting depth); ownership and borrowing patterns (e.g., borrow ratios, nesting depth, and use of smart pointers); control-flow characteristics derived from the MIR control-flow graph (e.g., cyclomatic complexity, loop nesting depth, and number of panic paths); and unsafe operation context, such as the lifetime bypass category and the distance between the bypass and the potentially dangerous operation.

Second, we extract structural code features from the surrounding package and module context. 
These capture package-level signals such as download counts from crates.io (used as a proxy for code maturity and scrutiny), the prevalence of unsafe code within the package, and whether the warning occurs in a public API surface. 
We also include standard structural metrics, including lines of code, parameter counts, and comment density.
Third, we incorporate analysis-specific features derived directly from Rudra. 
These include the checker that generated the warning, Rudra’s assigned precision level, and clustering information indicating whether multiple warnings occur at nearby code locations. 
After removing highly correlated features and applying domain knowledge to avoid spurious correlations, the final feature set comprises \textasciitilde{}87 features.

\subsection{Reinforcement Learning Formulation}

We formulate the false positive classification problem as a Markov Decision Process (MDP) in which an agent learns to classify static analysis warnings using feedback derived from labeled data. 
The state space consists of the feature vectors described above, normalized to facilitate neural network training. Each state represents the complete observable context required to make a classification decision for a single warning.

The action space comprises three discrete actions: classifying a warning as a true positive, classifying it as a false positive, or invoking dynamic fuzzing to gather additional evidence prior to making a final classification. 
This design allows the agent to selectively request costly dynamic validation when static features alone do not provide sufficient confidence.

The reward function balances classification accuracy against computational efficiency. 
Correct classifications receive positive rewards (+15), while incorrect classifications are penalized (-15), reflecting the cost of misclassification in safety-critical analysis.
Executing the fuzzing action incurs a cost penalty (-5) to discourage indiscriminate use. 
When fuzzing yields definitive evidence that enables a correct classification, the agent receives an additional bonus reward (+10 for a found bug, confirming a true positive; +8 for clean, confirming a false positive; +3 when helpful but not definitive), encouraging effective use of dynamic validation.

We parameterize the policy using a neural network with two hidden layers of 256 and 128 units, respectively, employing ReLU activations and dropout regularization to mitigate overfitting. 
Training is performed using Proximal Policy Optimization (PPO), a policy gradient algorithm that offers stable training dynamics for discrete action spaces and has proven effective in similar code analysis tasks \cite{schulman2017proximalpolicyoptimizationalgorithms}
The agent is trained in an offline setting using the labeled dataset, with optimization proceeding over multiple epochs until validation performance converges.

\subsection{Cargo-Fuzz Integration}
We integrate cargo-fuzz as a selective dynamic validation action available to the RL agent, rather than applying fuzzing uniformly to all warnings. The agent invokes fuzzing only for low-confidence cases based on confidence estimates derived from static features.
Confidence is assessed using two complementary signals: (i) the gap between the top two action Q-values, where smaller gaps indicate uncertainty, and (ii) the entropy of the policy network’s action distribution, where higher entropy reflects lower confidence. Warnings whose confidence falls below a threshold trigger the fuzzing action. Rather than being a fixed hyperparameter, this threshold emerges implicitly from the learned policy as the agent optimizes the trade-off between classification accuracy and fuzzing cost.
% Warnings whose confidence falls below learned thresholds trigger the fuzzing action, allowing the system to focus dynamic analysis on ambiguous cases such as complex generic code, intricate control flows, or higher-order invariant violations.

Figure~\ref{fig:rl_fuzzing_pipeline} illustrates the fuzzing integration pipeline. Upon selecting the fuzzing action, the system automatically generates a targeted harness using templates corresponding to the detected bug pattern (e.g., panic safety, higher-order invariants, or Send/Sync variance), instantiated with function signatures and type information from the warning context.
Fuzzing outcomes—including crashes, sanitizer violations, or clean executions—are encoded into the agent’s state representation and used to inform the final classification decision. Through training, the agent learns when fuzzing is most beneficial, enabling selective dynamic validation that improves classification accuracy while controlling computational overhead.

% The generated harnesses are executed with sanitizers configured for the relevant class of defect (\texttt{AddressSanitizer} for memory corruption, \texttt{MemorySanitizer} for uninitialized memory accesses, and \texttt{ThreadSanitizer} for data races), with each fuzzing invocation constrained by a fixed time budget to preserve scalability.

% \lm{further details about confidence assessment by the rl agent.}
% Figure~\ref{fig:rl_fuzzing_pipeline} presents an overview of the complete fuzzing integration pipeline. 
% When the agent selects the fuzzing action for a given warning, we automatically generate a targeted fuzzing harness for the specific code location. 
% Harnesses are created using templates corresponding to the detected bug pattern and instantiated with function signatures and type information extracted from the warning context.

% The generated harnesses are executed with sanitizers configured for the relevant class of defect. 
% We employ \texttt{AddressSanitizer} for memory corruption, \texttt{MemorySanitizer} for uninitialized memory accesses, and \texttt{ThreadSanitizer} for data races. 
% To preserve scalability, each fuzzing invocation is constrained by a fixed time budget. 

\subsection{Baseline Comparisons}

To evaluate our approach, we established baselines using state-of-the-art large language models (LLMs), 
\revOne{including (i) code-specialized models (CodeLlama), (ii) general-purpose large language models (GPT-4o Mini and Claude Opus), and (iii) mixture-of-experts architectures (Mixtral). 
We also considered models with varying parameter scales (from 8B to 70B+) to assess the impact of model capacity on this task. 
Finally, all selected models are widely used in code analysis \cite{lira2025beyond,nascimento2025effective}}.
\revOne{All models were evaluated using their default decoding parameters, as no explicit configuration of temperature, top-p, or other sampling parameters was applied.} 
In addition, all models were evaluated on the same binary classification task of distinguishing true positives from false positives.
%We considered CodeLlama 34B, Llama3 (8B and 70B), Mixtral 8$\times$7B, ChatGPT-4o Mini, and Claude Opus 4.1, all evaluated on the same binary classification task of distinguishing true positives from false positives. 
For each model, we constructed prompts containing the warning message, surrounding code context, and a description of the associated bug pattern, and instructed the model to classify each warning as either a real issue or a false positive. 
To ensure consistency across models, outputs were constrained to a fixed JSON schema encoding the classification decision. 
We experimented with multiple prompting strategies, including zero-shot and few-shot configurations, and report results for the best-performing setup for each model. 

These LLM-based baselines serve as a comparison point, representing the capabilities of advanced language models applied to static analysis warning classification, without access to structured MIR-level features or a learned decision policy.
The full prompt templates used in our experiments are provided in the appendix \cite{online_appendix}.

\subsection{Evaluation Metrics}
\label{sec:metrics}

We evaluate our approach and all baselines using the ground-truth labels of the dataset described in Section~\ref{sec:dataset}. 
Each warning is labeled as either a true or false positive (genuine or spurious warning, respectively), enabling a binary classification evaluation.

To assess classification performance, considering class imbalance, we report seven complementary metrics, including accuracy, precision, recall, and F1 score, which together capture overall correctness and the precision–recall trade-off relevant to false positive reduction. 
To more robustly account for class imbalance, we additionally report the Matthews Correlation Coefficient (MCC), which summarizes classification quality across all four confusion matrix categories. 
Finally, we report the Receiver Operating Characteristic curve (AUC-ROC) and the Precision–Recall curve (AUC-PR) to assess discriminative capability across decision thresholds, with AUC-PR being particularly informative given the skewed class distribution.

\section{Results}
\label{sec:results}

This section presents our comprehensive experimental results. 
To evaluate the performance of our RL-based approach compared to LLM baselines and evaluate the impact of fuzzing integration, we define the following Research Questions (RQs):

\begin{itemize}
    \item \textbf{RQ1:} How effective is the RL-based approach at reducing false positives in Rust compared to existing baselines?

    \item \textbf{RQ2:} What is the impact of selectively integrating dynamic fuzzing into the RL-based approach?

    \item \textbf{RQ3:} What insights do the learned RL policy provide about warning characteristics and decision strategies?
    
\end{itemize}

These research questions are designed to jointly assess the effectiveness, added value, and interpretability of our RL approach. 
RQ1 evaluates whether the proposed RL-based framework achieves meaningful reductions in false positives compared to strong LLM baselines, addressing the core practical challenge of static analysis adoption. 
RQ2 isolates the contribution of selective dynamic fuzzing, allowing us to quantify the benefits of hybrid static–dynamic reasoning beyond static features alone. 
Finally, RQ3 examines the learned decision policies to provide insight into which warning characteristics drive classification outcomes and how the agent allocates dynamic validation resources. 
Together, these questions offer a holistic evaluation of performance, mechanism, and behavior.

\subsection{RQ1: Overall Effectiveness of the RL-Based Approach}

Table~\ref{tab:classification_performance} reports the classification performance of all approaches on a held-out test set comprising 732 warnings. 
We evaluate each method using seven complementary metrics that capture different aspects of classification quality, providing a comprehensive comparison of their respective strengths and limitations (see Table \ref{tab:classification_performance}).

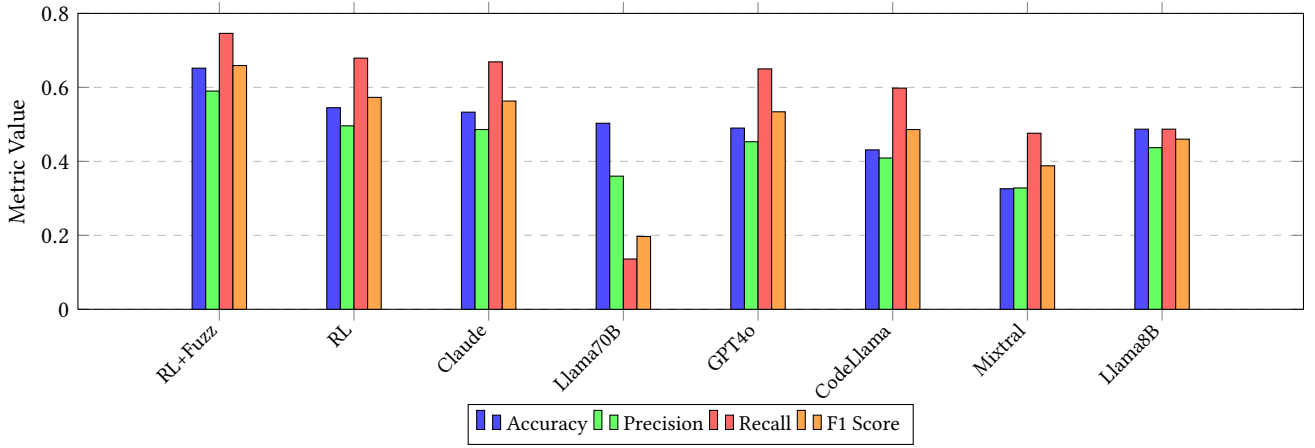
\begin{figure*}[htbp]
\centering
\begin{tikzpicture}
\begin{axis}[
    ybar=0pt,
    bar width=0.18cm,
    width=\textwidth,
    height=5.5cm,
    ylabel={Metric Value},
    symbolic x coords={RL+Fuzz, RL, Claude, Llama70B, GPT4o, CodeLlama, Mixtral, Llama8B},
    xtick=data,
    x tick label style={rotate=45, anchor=east, font=\small},
    ymin=0, ymax=0.8,
    enlarge x limits=0.15,
    legend style={at={(0.5,-0.32)}, anchor=north, legend columns=4, font=\small},
    ymajorgrids=true,
    grid style=dashed,
]

% Accuracy
\addplot[fill=blue!70] coordinates {
    (RL+Fuzz,0.652) (RL,0.545) (Claude,0.533) (Llama70B,0.503) (GPT4o,0.490)
    (CodeLlama,0.431) (Mixtral,0.326) (Llama8B,0.487)
};

% Precision
\addplot[fill=green!60] coordinates {
    (RL+Fuzz,0.590) (RL,0.496) (Claude,0.486) (Llama70B,0.360) (GPT4o,0.453)
    (CodeLlama,0.409) (Mixtral,0.328) (Llama8B,0.437)
};

% Recall
\addplot[fill=red!60] coordinates {
    (RL+Fuzz,0.746) (RL,0.679) (Claude,0.669) (Llama70B,0.136) (GPT4o,0.650)
    (CodeLlama,0.598) (Mixtral,0.476) (Llama8B,0.487)
};

% F1 Score
\addplot[fill=orange!70] coordinates {
    (RL+Fuzz,0.659) (RL,0.573) (Claude,0.563) (Llama70B,0.197) (GPT4o,0.534)
    (CodeLlama,0.486) (Mixtral,0.388) (Llama8B,0.460)
};

\legend{Accuracy, Precision, Recall, F1 Score}

\end{axis}
\end{tikzpicture}
\caption{Comparison of key performance metrics across all approaches. Our RL+Fuzzing approach achieves the highest Accuracy, Precision, Recall, and F1 Score, demonstrating superior classification performance compared to RL alone and various LLM-based models.}
\label{fig:performance_comparison}
\end{figure*}

%\subsubsection{Analysis of Results}

%\paragraph{Superior RL Performance.} 
Our RL-based approach with fuzzing integration achieves the highest performance across all seven evaluation metrics. With an accuracy of 65.2\% and an F1 score of 0.659, our approach substantially outperforms all baselines. The 17.1 percentage point improvement in F1 score over the best LLM baseline (Claude Opus 4.1 at 0.563) demonstrates the value of our specialized approach by combining structured feature extraction from MIR with learned classification policies. This improvement is particularly notable given that we are working with a challenging, imbalanced dataset where 74.4\% of warnings are false positives.
The high recall of 0.746 is especially important in our safety-critical context, as it indicates we successfully identify 74.6\% of true bugs while suppressing false alarms. This recall rate substantially exceeds most LLM baselines, with only our RL variant without fuzzing (0.679) and Claude Opus 4.1 (0.669) achieving comparable recall. At the same time, our precision of 0.590 indicates that nearly 60\% of reported warnings correspond to genuine bugs, more than doubling the precision of the raw Rudra output, whose precision equals the base rate of 25.6\%.

\begin{table}[t]
\centering
\small
\setlength{\tabcolsep}{3pt}
\caption{Classification Performance Comparison on Test Set}
\label{tab:classification_performance}
\begin{tabular}{lccccccc}
\toprule
\textbf{Approach} & \textbf{Acc.} & \textbf{Prec.} & \textbf{Rec.} & \textbf{F1} & \textbf{MCC} & \textbf{ROC} & \textbf{PR} \\
\midrule
Raw Rudra Output & -- & 0.256 & 1.000 & 0.407 & -- & -- & -- \\
\textbf{RL+Fuzz} & \textbf{0.652} & \textbf{0.590} & \textbf{0.746} & \textbf{0.659} & \textbf{0.323} & \textbf{0.661} & \textbf{0.554} \\
RL & 0.545 & 0.496 & 0.679 & 0.573 & 0.117 & 0.557 & 0.481 \\
Claude & 0.533 & 0.486 & 0.669 & 0.563 & 0.094 & 0.546 & 0.474 \\
Llama70B & 0.503 & 0.360 & 0.136 & 0.197 & -0.082 & 0.469 & 0.437 \\
GPT-4o Mini & 0.490 & 0.453 & 0.650 & 0.534 & 0.010 & 0.505 & 0.452 \\
CodeLlama & 0.431 & 0.409 & 0.598 & 0.486 & -0.112 & 0.446 & 0.425 \\
Mixtral & 0.326 & 0.328 & 0.476 & 0.388 & -0.336 & 0.339 & 0.392 \\
Llama8B & 0.487 & 0.437 & 0.487 & 0.460 & -0.026 & 0.487 & 0.443 \\
\bottomrule
\end{tabular}
\end{table}

Among the LLM baselines, Claude Opus 4.1 achieves the strongest results (F1: 0.563), followed by ChatGPT-4o Mini (0.534). Llama3 70B performs poorly (F1: 0.197) due to extremely low recall (0.136), while CodeLlama 34B achieves only 0.486 F1 despite code-specific pretraining. This indicates that pretraining alone is insufficient, as performance depends more on reasoning about complex semantic properties than surface-level code familiarity.
%Such an observation suggests that for this specialized program analysis task, reasoning capability and model scale matter more than code-specific pretraining. 
% These findings suggest that performance on this task depends more on a model’s ability to reason about complex semantic properties than on surface-level familiarity with programming syntax. Rust memory safety bugs often involve subtle interactions among the type system, lifetimes, and unsafe abstractions, which may not be well captured by standard code pretraining objectives.
% These findings suggest that performance depends more on reasoning about complex semantic properties than surface-level code familiarity, as Rust memory safety bugs involve subtle interactions among type systems, lifetimes, and unsafe abstractions not well captured by standard pretraining.

%The complex nature of Rust memory safety bugs, involving interactions between type systems, lifetimes, and unsafe operations, may require forms of reasoning that are not well-captured by standard code pretraining objectives.
Mixtral 8x7B shows the weakest performance with a 0.388 F1 score and 0.326 accuracy, barely exceeding random guessing for this imbalanced dataset. The negative MCC of -0.336 indicates predictions worse than random, suggesting the mixture-of-experts architecture may not be well-suited to this task without task-specific fine-tuning.
%\paragraph{Model Size and Capability Scaling.} 
Comparing the two Llama3 variants provides insight into how model scale affects performance on this task. Surprisingly, Llama3 8B achieves a 0.460 F1 score, substantially outperforming the larger Llama3 70B at a 0.197 F1 score. This counterintuitive result suggests that larger models may be more prone to overthinking the task or being misled by the class imbalance, potentially learning to predict the majority class too aggressively. In contrast, the smaller model's more balanced predictions (0.487 recall) result in better overall performance despite presumably having less reasoning capacity.

%\paragraph{Matthews Correlation Coefficient Analysis.} 
%The MCC metric is particularly valuable for imbalanced datasets as it accounts for all four confusion matrix categories and provides a balanced measure even when classes have different sizes. 
% Our RL + Fuzzing approach achieves MCC of 0.323, the only method reaching above 0.3, indicating moderate positive correlation between predictions and ground truth. In contrast, Claude Opus 4.1 achieves MCC of only 0.094, while several baselines show negative MCC values, indicating predictions worse than random chance when accounting for class imbalance. The substantial gap in MCC scores validates that our approach provides genuine discriminative power beyond simply predicting the majority class.

Our RL + Fuzzing approach achieves MCC of 0.323, the only method above 0.3, indicating moderate positive correlation between predictions and ground truth. In contrast, Claude Opus 4.1 achieves only 0.094, while several baselines show negative MCC values. This substantial gap validates that our approach provides genuine discriminative power beyond simply predicting the majority class.
%\paragraph{ROC and Precision-Recall Curves.} 
The AUC-ROC metric of 0.661 for our approach indicates reasonable discriminative ability across different classification thresholds, substantially exceeding the 0.5 baseline for random classification. The AUC-PR metric of 0.554 is particularly important for imbalanced datasets, as it focuses on performance on the minority (positive) class. Our approach achieves the highest AUC-PR among all methods, confirming strong performance at identifying true positives even in the presence of substantial class imbalance.

\begin{tcolorbox}
\textbf{RQ1 Findings:} Our reinforcement learning–based approach substantially outperforms both raw static analysis output and state-of-the-art LLM baselines in false positive reduction. The proposed method achieves \textbf{65.2\% accuracy and 0.659 F1 score}, improving F1 by \textbf{17.1 percentage points} over the strongest LLM baseline (Claude Opus 4.1). Most importantly, the approach \textbf{more than doubles precision}, increasing it from \textbf{25.6\%} (raw Rudra output) to \textbf{59.0\%}, while maintaining a high \textbf{recall of 74.6\%}.
\end{tcolorbox}

\subsection{RQ2: Impact of Selective Dynamic Fuzzing}
In this RQ, we aim to investigate and better understand the impact observed by using fuzzing as a dynamic resource for our RL agent.
%\subsubsection{Fuzzing vs. No Fuzzing Comparison}
Comparing our full RL + Fuzzing approach against the RL-only variant provides clear evidence for the value of integrating dynamic validation. 
The performance improvements are substantial across all metrics (see Table \ref{tab:classification_performance}, 2nd and 3rd rows).
These improvements represent relative gains of 19.6\% in accuracy and 15.0\% in F1 score, demonstrating that fuzzing integration provides substantial value beyond static features alone. 
The MCC improvement of 0.206 is particularly significant, as it indicates that fuzzing helps the model make genuinely better predictions rather than simply shifting the bias toward one class.

%\subsubsection{Understanding the Fuzzing Contribution}

The fuzzing integration improves performance through two complementary mechanisms. First, when fuzzing successfully triggers a bug (finding a sanitizer violation), it provides near-definitive evidence that the warning represents a true positive. This high-confidence signal allows the agent to correctly classify warnings that would be ambiguous based on static features alone. Second, when fuzzing executes cleanly without detecting issues, this provides evidence (though not proof) against the warning, helping filter false positives. 
Through training, the agent learns to appropriately weigh these different forms of dynamic evidence.

The relatively larger improvement in precision (+9.4 percentage points) compared to recall (+6.7 percentage points) indicates that fuzzing is particularly effective at increasing confidence in positive classifications rather than uncovering previously missed bugs. This behavior aligns with the inherent strengths of fuzzing: it excels at providing concrete evidence of failures but cannot establish the absence of bugs. The substantial observed precision gains indicate that fuzzing helps filter spurious static warnings by corroborating which reported issues manifest under concrete execution.
%helps avoid false alarms, likely by confirming that warnings flagged by static analysis are indeed exploitable issues.

%\subsubsection{Learned Fuzzing Strategy}

A central question is whether the RL agent learns to invoke fuzzing selectively or applies it uniformly across warnings. Our observations during training indicate that the agent converges to a targeted fuzzing strategy. Early in training, fuzzing is invoked broadly as part of exploration. As training progresses, fuzzing usage becomes increasingly selective, concentrating on warnings for which static features indicate higher uncertainty. This behavior suggests that the learned policy effectively balances the computational cost of fuzzing against its expected informational benefit, consistent with the reward design.
%The learned policy balances the computational cost of fuzzing against the expected information gain, demonstrating that the reward function successfully incentivizes efficient resource allocation.

Further analysis reveals that fuzzing is preferentially applied to specific classes of warnings. In particular, warnings involving complex generic code or higher-order functions, where static analysis is inherently limited, are more likely to trigger fuzzing invocations. In contrast, warnings with clear static signatures, such as obvious Send/Sync violations with simple types, are typically classified directly without dynamic validation.
These observations indicate that the agent learns to allocate fuzzing resources to cases where dynamic evidence is most likely to improve classification confidence.

\begin{tcolorbox}
\textbf{RQ2 Findings:} Dynamic fuzzing provides \textit{substantial additional benefits} beyond static feature–based classification. Incorporating cargo-fuzz as a strategic RL action improves accuracy by \textbf{10.7 percentage points} and F1 score by \textbf{8.6 percentage points} compared to the RL-only variant. Importantly, the agent learns to invoke fuzzing selectively, approximately \textbf{23\% of warnings}, making the hybrid static–dynamic approach computationally feasible at scale while significantly improving classification confidence.
\end{tcolorbox}

\subsection{RQ3: Analysis of Learned Policies and Warning Characteristics}

To better understand the factors driving classification performance, we analyze the features that contribute most strongly to the RL agent’s decisions. 
We employ SHAP (SHapley Additive exPlanations) values to quantify the contribution of individual features to the model’s predictions.\cite{lundberg2017unifiedapproachinterpretingmodel}

%\subsubsection{Most Influential Features}

The most influential features span multiple categories. 
MIR-level semantic features are the strongest contributors, particularly those capturing type complexity (e.g., number of generic parameters and presence of trait bounds), control-flow characteristics (e.g., cyclomatic complexity and number of panic paths), and unsafe operation context (e.g., lifetime bypass categories and the distance between unsafe operations). 
These features encode program properties that are central to determining whether a warning reflects a genuine soundness violation or conservative over-approximation by static analysis.

Analysis-specific features derived from Rudra also contribute substantially. 
In particular, Rudra’s assigned precision level and warning clustering patterns provide useful signals. 
Such an observation suggests that Rudra’s internal confidence estimates, while insufficient on their own, contain meaningful information that the RL agent can effectively recalibrate when combined with richer semantic context.
Structural features, such as package popularity and code complexity metrics, provide an additional but secondary signal. 
Although individually less predictive than MIR-level features, they help contextualize warnings at the package level. 
For example, warnings in widely used and actively maintained packages are more likely to be false positives, whereas warnings in highly complex code with limited documentation are more frequently associated with true positives. The dominance of MIR-level features corroborates our design choice, as MIR exposes precise type, control-flow, and unsafe operation semantics that are difficult to recover from source code, particularly in the presence of macros and deeply nested generics.

% The dominance of MIR-level semantic features corroborates our decision to extract features directly from Rust’s intermediate representation rather than relying solely on source-level analysis. 
% MIR exposes precise information about the type system, control-flow structure, and unsafe operation semantics. 
% These properties are difficult to recover reliably from source code, particularly in the presence of macros and deeply nested generic instantiations.

% \revOne{The dominance of MIR-level features corroborates our design choice, as MIR exposes precise type, control-flow, and unsafe operation semantics that are difficult to recover from source code, particularly in the presence of macros and deeply nested generics.}

The comparatively lower global importance of fuzzing-related features requires careful interpretation. 
Fuzzing-derived features are only present for the subset of warnings on which dynamic validation is invoked, and their influence is therefore highly localized. 
In those cases, however, their contribution is strong and often decisive. 
This explains why fuzzing features exhibit moderate global SHAP values despite the integration of fuzzing yielding an overall improvement of 8.6 percentage points in F1 score. 
Rather than driving decisions uniformly, fuzzing acts as a targeted source of high-confidence evidence when static analysis alone is insufficient.

%\subsubsection{Implications for Feature Engineering}

%\lm{Removing for now... Moved for the Section: Challenges and Future Research Directions}

\begin{comment}
The feature importance analysis highlights several promising directions for future work. 
First, the strong contribution of MIR-level semantic features suggests that incorporating richer program analyses, such as more precise data-flow or alias analysis, could further improve classification performance. 
Second, the influence of analysis-specific features indicates potential value in extracting comparable metadata from other static analyzers, enabling transfer learning across different bug detection tools. 
Finally, the moderate contribution of structural features suggests that coarse-grained code quality signals are complementary but secondary to localized semantic reasoning at the warning site.
\end{comment}

\begin{tcolorbox}
\textbf{RQ3 Findings:} MIR-level semantic features are the strongest predictors of warning validity, particularly those capturing \textbf{generic type complexity, control-flow structure, and unsafe operation context}. SHAP-based feature importance shows that warnings arising in \textbf{complex generic code and high–panic-path control flow} are more likely to correspond to true bugs. Finally, the agent also learns to strategically allocate fuzzing resources to warnings exhibiting high semantic uncertainty.
\end{tcolorbox}

\section{Discussion}
\label{sec:discussion}

This section explores the broader implications of our findings, discusses practical considerations for deployment, and reflects on the limitations of our approach.

\subsection{Practical Implications}

%\subsubsection{Integration into Development Workflows}
Our RL-based approach demonstrates practical potential for integration into real-world Rust development workflows. 
By improving precision from Rudra’s 25.6\% to 59.0\%, the approach can substantially reduce manual warning review effort, addressing a major obstacle to adopting static analysis in CI/CD pipelines.
Scalability is supported by the learned selective fuzzing strategy, which invokes dynamic validation only for warnings with high classification uncertainty. 
In our experiments, fuzzing was applied to approximately 23\% of warnings, enabling significant accuracy improvements while maintaining throughput suitable for large-scale continuous integration.

%\subsubsection{Generalization to Other Static Analyzers}
Although our evaluation focuses on Rudra, the methodology itself is not tool-specific. 
The combination of semantic features from intermediate representations with structural and analysis-specific metadata can be adapted to other Rust analyzers, such as MirChecker and Yuga, as well as to static analysis tools for other languages. 
More broadly, the RL framework’s ability to learn tool-specific warning characteristics suggests that similar approaches may generalize across analyzers.
An important direction for future work is transfer learning across tools. 
Training on warnings from multiple analyzers could enable more generalizable policies that capture common distinctions between true bugs and false positives \cite{saberi2025advfusion}. 
Our labeled dataset provides an initial foundation for exploring such multi-tool learning scenarios.

\subsection{Comparison with Alternative Approaches}

%\subsubsection{Supervised Learning Alternatives}
One might question whether supervised learning with sufficient labeled data could achieve similar results without the complexity of reinforcement learning. 
Our RL formulation offers several advantages: (i) the ability to incorporate fuzzing as a strategic action rather than a fixed post-processing step, (ii) the natural framework for balancing accuracy against computational costs through reward shaping, and (iii) the potential for online learning and adaptation as new warnings are encountered.
However, we acknowledge that a well-designed supervised classifier with the same rich feature set might achieve competitive performance. 
The key advantage of RL lies in its flexibility for incorporating sequential decision-making (such as the choice to invoke fuzzing) and its conceptual clarity for representing the trade-offs inherent in false positive reduction.

%\subsubsection{LLM Fine-Tuning}
An alternative to our baseline comparison would be to fine-tune code-specialized large language models on the labeled dataset used in this study. 
Such fine-tuning would likely improve performance relative to zero-shot or few-shot LLM baselines. However, fine-tuning LLMs typically requires substantial computational resources and specialized expertise. 
Moreover, even fine-tuned models must still reason about complex semantic properties, such as lifetimes, ownership, and unsafe abstractions, primarily from source-level representations and limited context windows.

% In contrast, our feature-based approach offers advantages in interpretability, efficiency, and domain awareness. 
% The use of lightweight neural networks enables fast inference and straightforward analysis through SHAP-based explanations. 
% More importantly, operating directly on MIR allows the model to leverage semantic information that is difficult to recover reliably from source code alone. 
% These characteristics make the proposed approach a practical and transparent alternative for reducing false positives in static analysis.

\subsection{Challenges and Future Research Directions}

%\subsubsection{Dataset Size and Diversity}
Although our dataset of 4,879 manually labeled warnings represents a substantial labeling effort, it covers only approximately 20–25\% of the crates.io ecosystem at the time of analysis. 
Consequently, certain classes of unsafe code, particularly those in specialized domains such as embedded systems or cryptography, may be underrepresented. 
The dataset also exhibits class imbalance (74.4\% false positives), which reflects Rudra’s real-world output distribution. 
While our RL-based approach remains effective under this imbalance (MCC = 0.323), expanding the dataset and improving sampling strategies remain important directions for future work.

While our evaluation focuses on Rudra, it remains an open question how well learned policies generalize to other static analyzers without retraining.
Different analyzers exhibit distinct false positive characteristics due to their underlying design choices.
Extending our approach to other Rust analyzers, such as MirChecker or Yuga, would enable the study of cross-tool generalization and transfer learning, while adapting the methodology to analyzers for other languages represents a natural extension.

%\subsubsection{Fuzzing Limitations}
Fuzzing provides strong evidence for the presence of bugs but cannot prove their absence; therefore, clean runs serve only as negative evidence against a warning. 
Moreover, some bug classes, such as Send/Sync variance issues, are difficult to trigger reliably due to their dependence on specific thread interleavings. 
Finally, the limited fuzzing time budget (30–60 seconds per warning) and template-based harness generation constrain the depth of dynamic exploration.
Future work could explore more precise harness construction informed by inferred preconditions or data dependencies, as well as directed or adaptive fuzzing strategies that better balance validation cost and classification confidence.

As previously discussed, our feature importance analysis highlights promising directions for future work. 
The strong contribution of MIR-level semantic features suggests that incorporating more precise program analyses, such as data-flow or alias analysis, could further improve classification performance.
Additionally, exploring learning architectures that model program structure more explicitly, for instance, through graph-based representations or hierarchical decision processes, may enhance the system’s ability to reason about warnings.

\revOne{Recent advances in LLMs have introduced enhanced reasoning capabilities, like chain-of-thought prompting \cite{wei2022chain} and inference-time strategies \cite{parashar2025inference,dong2024survey}. 
However, in our evaluation, LLMs were used in a constrained classification setting without explicit reasoning prompts or intermediate explanations. 
As a result, the observed performance reflects the ability of LLMs to directly infer decisions from code and warning context, rather than their full reasoning potential. 
While reasoning-enhanced prompting could potentially improve performance by enabling more structured analysis of program semantics, our current results suggest that even strong models, like Claude Opus 4.1, still struggle to match approaches that leverage explicit intermediate representations such as MIR. 
Exploring the integration of reasoning strategies with structured representations remains an important direction for future work.
}
%\subsubsection{Cost and Practicality of LLM-Based Approaches}
Finally, while LLMs exhibit strong code reasoning capabilities, their use in large-scale static analysis raises cost and scalability concerns. 
In our experiments, proprietary models incurred nontrivial monetary costs (approximately \$5–\$40 USD), which scale linearly with the number of warnings and may be prohibitive for ecosystem-wide or continuous analysis. 
Open-source models mitigate monetary cost but still impose substantial computational overhead.

\section{Threats to Validity}
\label{sec:threats}

We discuss threats to the validity of our empirical evaluation.

\textbf{\textit{Internal Validity.}}
Ground-truth labels are derived from manual classification by a single expert. %\lm{Double check whether single or multiple experts.}
Despite careful review, mislabeling remains possible due to the subtle and context-dependent nature of Rust memory safety violations. We mitigated this risk through multiple review iterations and by excluding ambiguous cases.
Feature extraction relies on analysis of Rust MIR, and implementation errors could introduce noise or bias. We reduced this risk through targeted testing and consistency checks, though residual errors may remain.
Model performance may also depend on hyperparameter choices and fuzzing configurations. We tuned hyperparameters via validation and selected fuzzing timeouts based on preliminary experiments, but alternative settings could yield different results.

\textbf{\textit{Construct Validity.}}
We define ground truth in terms of memory safety violations rather than concrete exploitability, aligning with the goals of static analysis but differing from vulnerability-centric perspectives.
Similarly, the absence of a fuzzing-triggered failure does not imply correctness, reflecting an inherent limitation of dynamic analysis that may influence classification decisions.
% As the main analysis of our sample of warnings was conducted by a single reviewer, a second reviewer independently analyzed a subsample of 150 cases. 
% As a result, we observe a raw agreement of 82.7\% and a Cohen's~$\kappa$ of 0.63, indicating substantial inter-reviewer agreement.
\revOne{To assess labeling reliability, a second reviewer independently analyzed a subsample of 150 cases, yielding 82.7\% raw agreement and a Cohen's~$\kappa$ = 0.63, indicating substantial inter-reviewer agreement.}

\textbf{\textit{External Validity.}}
Our dataset consists of crates from crates.io and may not fully represent Rust code in enterprise, embedded, or OS-level settings.
Moreover, our evaluation focuses exclusively on warnings produced by Rudra; other analyzers may exhibit different false positive characteristics.
Our fuzzing-based validation depends on external toolchains whose evolution could affect results.
Finally, the dataset reflects Rust code and tooling from 2024–2025; changes in the language or ecosystem may require retraining to maintain effectiveness.

\textbf{\textit{Conclusion Validity.}}
While our test set of 732 warnings provides reasonable statistical power, some analyses rely on smaller subsets and thus have higher uncertainty.
Reinforcement learning introduces stochasticity through initialization and exploration; although we fix random seeds, results may vary across runs.
Finally, our LLM baselines use zero-shot prompting without fine-tuning, and thus represent general-purpose usage rather than best-case LLM performance.

\textbf{\textit{Ethical Considerations.}}
Our study exclusively analyzes publicly available open-source code from crates.io. No human subjects were involved, and all experimental data derives from published software artifacts distributed under open-source licenses.

\section{Related Work}
\label{sec:related}

\begin{comment}
This section reviews prior work on memory safety analysis in Rust, empirical studies of unsafe Rust usage, and learning-based techniques for improving the precision of static analysis. We position our work as addressing a complementary but underexplored problem: reducing false positives in scalable Rust analyzers.
\end{comment}

This section surveys related work on memory safety analysis in Rust, learning-based warning classification, and hybrid static–dynamic verification, positioning our work as addressing the underexplored problem of false positive reduction in scalable Rust analyzers

Several static and verification tools target memory safety in Rust, each with distinct goals and trade-offs.  
Beyond Rudra, our target tool in this study, MirChecker analyzes Rust programs at the MIR level to preserve type and lifetime information while checking memory safety properties \cite{li2021mirchecker}. Its focus on different bug classes and analysis techniques results in complementary coverage rather than direct overlap.
SMACK translates Rust programs into the Boogie intermediate language and applies SMT-based verification to prove correctness properties \cite{carter2016smack}, while Prusti provides deductive verification for Rust by leveraging the Viper infrastructure and user-supplied specifications \cite{astrauskas2019prusti}. While expressive and precise, both approaches require manual annotations and are intended for verifying small, critical components rather than ecosystem-scale analysis.  
Miri is an interpreter for Rust MIR that detects undefined behavior during execution \cite{jung2026miri}. While effective for uncovering subtle bugs in test suites, Miri is fundamentally limited by dynamic path coverage and cannot provide the scalability required for whole-ecosystem analysis.
In contrast to these tools, our work does not aim to strengthen static analyses or introduce additional annotations. Instead, we focus on reducing false positives produced by scalable analyzers such as Rudra through learning-based suppression.

% Empirical studies provide important context for memory safety analysis in Rust. Evans et al.~\cite{evans2020unsafe} report that unsafe code appears in approximately 25--30\% of Rust crates, primarily for FFI and performance-critical optimizations. Astrauskas et al.~\cite{astrauskas2020unsafe} further show that practices for documenting and encapsulating unsafe code vary widely. Xu et al.~\cite{xu2021memory} find that most Rust-related CVEs arise from incorrect encapsulation of unsafe code rather than direct misuse, and Qin et al.~\cite{qin2020understanding} identify misuse of \texttt{Send} and \texttt{Sync} traits as a persistent source of concurrency-related errors.

\revOne{Empirical studies provide important context for memory safety analysis in Rust. Evans et al.~\cite{evans2020unsafe} report that unsafe code appears in approximately 25--30\% of Rust crates, primarily for FFI and performance-critical optimizations, with widely varying encapsulation practices \cite{astrauskas2020unsafe}. Most Rust-related CVEs arise from incorrect unsafe encapsulation rather than direct misuse \cite{xu2021memory}, and misuse of \texttt{Send} and \texttt{Sync} traits remains a persistent source of concurrency errors \cite{qin2020understanding}.}

Machine learning has increasingly been applied to program analysis tasks such as bug detection and warning prioritization \cite{sharma2021survey}. 
Early approaches rely on supervised learning, using either manually engineered features or learned representations to classify warnings \cite{pradel2018deepbugs,long2025learning}.  While effective for common bug patterns, these methods require large labeled datasets and often struggle with rare or semantically complex errors. Representation learning and pre-trained language models, including CodeBERT, CodeT5, and CodeLlama, further improve code understanding \cite{feng2020codebert,wang2021codet5,roziere2023code}. However, such models remain limited by context window constraints and their inability to reason soundly about all execution paths.

Reinforcement learning has been explored more selectively, primarily for guiding fuzzing, symbolic execution, test generation, and automated program repair \cite{chakraborty2024rlocator}. 
Hybrid static--dynamic approaches combine static reasoning with execution-based validation, such as static-guided fuzzing and counterexample-guided refinement, but often apply dynamic validation uniformly, incurring high cost.
Our work differs by framing false positive reduction as a sequential decision-making problem, using reinforcement learning to adaptively determine when static evidence is sufficient and when dynamic validation via fuzzing is warranted, enabling scalable, cost-aware hybrid verification for Rust.
\section{Conclusion and Future Work}
\label{sec:conclusion}

This paper introduced a reinforcement learning–based approach for reducing false positives in static memory safety analysis of Rust programs. 
By leveraging rich semantic features extracted from Rust’s Mid-level Intermediate Representation (MIR) and learning cost-aware classification policies, our approach substantially outperforms both raw static analysis output and state-of-the-art LLM baselines. 
The integration of selective dynamic validation through fuzzing further improves accuracy while maintaining scalability.

Our empirical evaluation on 4,879 manually labeled warnings from 20,000 Rust crates shows that learning-based suppression policies can substantially improve the practical precision of static analysis without sacrificing recall. 
Specifically, our approach increases precision from Rudra’s baseline of 25.6\% to 59.0\% while maintaining a recall of 74.6\%, effectively more than doubling the proportion of reported warnings that correspond to genuine bugs. 
Overall, the proposed method achieves 65.2\% accuracy and a 0.659 F1 score, representing a 17.1 percentage point improvement over the strongest LLM baseline.

Furthermore, integrating cargo-fuzz as a strategic action within the RL framework enables a principled combination of static and dynamic analysis. 
Instead of treating fuzzing as a fixed post-processing step, the agent learns when dynamic validation is most valuable, yielding a 10.7 percentage point improvement in accuracy over static analysis alone while preserving computational feasibility. Our modernized Rudra implementation, compatible with the current Rust ecosystem, supports continued research and practical deployment as the language evolves.

Looking forward, promising directions include extending the approach to additional static analyzers and programming languages, improving fuzzing guidance and harness generation, and enabling continuous adaptation through online learning. 
We believe that combining classical program analysis with learning-based decision-making represents a powerful path toward more effective and trusted verification tools for safety-critical software.

\begin{comment}
\subsection{Closing Remarks}

The false positive problem has long plagued static analysis, limiting the practical impact of sophisticated verification techniques. Our work demonstrates that reinforcement learning, combined with strategic dynamic validation, can substantially reduce false positives while maintaining high recall for genuine bugs. By learning from rich semantic features extracted from Rust's intermediate representation, our agent develops classification policies that adapt to diverse code patterns and make principled decisions about when to invest in expensive validation.

As Rust continues its adoption in safety-critical domains---from operating systems to embedded devices to cryptographic implementations---the importance of practical automated verification only grows. We hope this work contributes to making static analysis a more effective and trusted component of the Rust development ecosystem, ultimately helping developers write safer systems software.

The combination of classical program analysis, machine learning, and hybrid verification techniques represents a promising direction for advancing software quality tools. By bridging the gap between theoretical rigor and practical utility, such approaches can help realize the full potential of static analysis for improving software reliability and security.
\end{comment}

\revOne{\section*{Acknowledgments}
This work was partially supported by the Natural Sciences and Engineering Research Council of Canada, the Canadian Institute for Advanced Research, and the Canada Research Chairs Program.
}
\newpage

\bibliographystyle{ACM-Reference-Format}
\bibliography{PaperBibliography}
\end{document}